\documentclass [a4paper, 11pt] {article}
\usepackage{amsmath}
\usepackage{amssymb}
\usepackage{amsfonts}
\usepackage{latexsym}
\usepackage{amsmath}
\usepackage{mathtools}
\usepackage{empheq}
\usepackage{graphicx}
\usepackage{caption}
\usepackage{subcaption}
\usepackage[toc,page]{appendix}
\usepackage{multicol}
\usepackage{color}

\begin{document}

\title{\textbf{Resonant absorption of kink magnetohydrodynamic waves by a magnetic
twist in coronal loops }}

\author{Zanyar Ebrahimi\thanks{E-mail: Z.Ebrahimi@uok.ac.ir} ,
Kayoomars Karami\thanks{E-mail: KKarami@uok.ac.ir}\\
\small{Department of Physics, University of Kurdistan, Pasdaran St.,
Sanandaj, Iran}}

\maketitle

\begin{abstract}
There is ample evidence of twisted magnetic structures in the solar corona. This motivates us to consider the magnetic twist as the
cause of Alfv\'{e}n frequency continuum in the coronal loops, which can
support the resonant absorption as a rapid damping mechanism for
the observed coronal kink magnetohydrodynamic (MHD) oscillations. We model a coronal loop with a straight cylindrical magnetic
flux tube which has constant but different densities in the interior and exterior regions. The magnetic field is assumed to be constant and aligned with the cylinder axis everywhere except a thin layer near the boundary of the flux tube which has an additional small magnetic field twist. Then, we investigate a number of possible instabilities that may arise in our model. In the thin tube thin boundary approximation, we derive the dispersion relation and solve it analytically to obtain the frequencies and damping rates of the fundamental $(l=1)$ and first/second overtone $(l=2,3)$ kink
$(m=1)$ MHD modes. We conclude that the resonant absorption by the
magnetic twist can justify the rapid damping of kink MHD waves
observed in coronal loops. Furthermore, the magnetic twist in the
inhomogeneous layer can cause deviations from $P_1/P_2=2$ and
$P_1/P_3=3$ which are comparable with the observations.
\end{abstract}



\section{Introduction}

The first identification of transverse oscillations of coronal loops
was reported by Aschwanden et al. (1999) and Nakariakov et al.
(1999) using the Transition Region and Coronal Explorer (TRACE)
observations of 14 July 1998 in the 171 {\AA} Fe IX emission lines.
Nakariakov et al. (1999) for a loop with length of
$(130\pm6)\times10^{3}$ km and width of $(2.0\pm0.36)\times10^3$ km
reported a spatial oscillation with period of
$4.27\pm0.13$ min and decay time of $14.5\pm2.7$ min. They suggested
resonance of the global mode as the cause of such fast damping. On 17 April 2002,
the vertical polarization of coronal loops oscillations with period
of 3.9 min and decay time of 11.9 min were identified by Wang \&
Solanki (2004) using the TRACE observations in the 195 {\AA}
Fe XII emission line. According to
Roberts et al. (1984), the goal of coronal seismology is to deduce
the properties of the solar corona using observed parameters of
oscillations and waves. For instance, Nakariakov \& Ofman
(2001) applied a new method for determination of the local
magnetic field strength base on the observed length, density and
frequency of an oscillating coronal loop. For reviews on
coronal seismology, see e.g. De Moortel (2005), De Moortel \&
Nakariakov (2012), Andries et al. (2009) and Ruderman \&
Erd\'{e}lyi (2009).

The theory of resonant absorption of MHD waves was first established by Ionson
(1978) as a conceivable mechanism for heating of the solar corona. Since then, many theoretical works have been done to develop
this theory (see e.g. Davila 1987; Sakurai et al. 1991a,b;
Goossens et al. 1992; Steinolfson \& Davila 1993). In this
mechanism, energy of the global mode oscillations is transferred to the local
Alfv\'{e}n perturbations within a resonance layer inside the loop. The necessary condition for
this process is a gradient of Alfv\'{e}n frequency in this layer
which varies between the interior and exterior Alfv\'{e}n
frequencies of the loop. For a good review on the resonant absorption
see also Goossens et al. (2011).

Heating of the coronal loops by the resonant absorption of MHD waves
was studied by Erd\'{e}lyi \& Goossens (1995). Solving the
visco-resistive MHD equations of motion, they concluded
that under coronal conditions both viscus and ohmic dissipation
mechanisms are important. Erd\'{e}lyi \& Goossens (1996) showed
that the equilibrium plasma flow in coronal flux tubes affects the
resonant absorption rate due to driving waves.

Ruderman \& Roberts (2002) investigated the resonant
absorption of kink mode oscillations in a coronal loop with
radial density inhomogeneity in the zero-beta approximation. They
concluded that the oscillations of coronal loops are coherent only in the presence of small scale inhomogeneities in density. Goossens et al. (2002) showed that the
damped quasi-modes give an accurate description of rapid damping of the
observed coronal loop oscillations if the length
scale of inhomogeneity changes from loop to loop. Also damping of quasi-modes is completely consistent with large estimates of the Reynolds numbers in the corona ($10^{14}$).

Van Doorselaere et al. (2004) investigated the oscillations of kink modes in
coronal loops by the
LEDA numerical code (van der Holst et al. 1999). Taking into account the large inhomogeneity length scales, Van Doorselaere et
al. (2004) showed that the rapid damping of oscillating coronal loops can be justified by resonant absorption without resorting to the Reynolds numbers smaller than the classical values. They concluded that the numerical results of damping rates can deviate from the analytical ones obtained in thin boundary approximation, by up to $25\%$ .

Terradas et al. (2006a) investigated the temporal evolution of resonant absorption
in a one-dimensional
cylindrical coronal loop. They found that when the coronal loop is
excited by the external perturbation, the first stage of the loop
oscillation has the leaky behavior. After that, the loop oscillates like the kink mode and then it is dissipated by the mechanism of resonant absorption.
Terradas et al. (2006b) developed their previous work and found that considering the curvature of coronal loops enhances the efficiency of resonant absorption slightly. They showed that there are two kink modes with polarizations mostly in the horizontal/vertical directions with respect to the photosphere. These
modes show both resonant and leaky behavior at a same time.

The effect of longitudinal density stratification on the resonant
absorption of MHD waves in coronal loops with radial density gradients has also been studied in the literature (see e.g. Andries
et al. 2005; Karami et al. 2009). Karami et al. (2009) showed that
in the zero-beta approximation, when the stratification parameter
increases, both the period and damping time of the kink and
fluting modes decrease but the stratification does not
affect the ratio of frequencies to damping rates. They
further showed that the ratio of fundamental period to first overtone one decreases from its canonical value $P_1/P_2=2$ when
the stratification parameter increases.

Besides the above considerations, there are
observational evidences for twisted magnetic fields in
coronal loops. Chae et al. (2000) reported the traces of rotational
motions in coronal loops and suggested that the existence of an
azimuthal magnetic field component that encircles the axis may be
required to guide the rotational motions. Chae \& Moon (2005)
considered a model of twisted flux tube in which the force of pressure gradient is balanced by the tension force of the azimuthal magnetic field component. They suggested that the constriction of plasma can be used to determine the magnetic twist in coronal loops.
Kwon \& Chae (2008) measured the magnetic twist of 14 coronal loops. The magnetic twist value $\phi_{\rm twist}$ of the loops was in the range $[0.22\pi,1.73\pi]$.

Many theoretical works have also been done on the effect of twisted
magnetic fields on the MHD waves in coronal loops (see e.g. Bennett
et al. 1999; Erd\'{e}lyi \& Carter 2006; Erd\'{e}lyi \& Fedun
2006, 2007, 2010; Carter \& Erd\'{e}lyi 2007, 2008; Ruderman 2007, 2015;
Karami \& Barin 2009; Karami \& Bahari 2010, 2012; Terradas \&
Goossens 2012). Ruderman (2007) considered a straight flux
tube in the zero-beta approximation with a magnetic twist inside the
loop proportional to the distance from the tube axis. Using the
asymptotic analysis in the limit of small twists, Ruderman
obtained an analytical solution for perturbations inside the loop
and showed that the magnetic twist does not affect the standing kink modes.
Karami \& Bahari (2012) extended the work of Ruderman (2007)
to a magnetically twisted flux tube containing both core and annulus regions. They showed that the frequencies and the
period ratio $P_1/P_2$ of the fundamental and first overtone
nonaxisymmetric kink and fluting modes are affected by the twist parameter of the annulus. Terradas \& Goossens (2012) studied the effect of magnetic twist on the kink oscillations of coronal
loops with a piecewise parabolic twist profile. They solved the MHD equations
using the PDE2D (Sewell 2005) code. Terradas \& Goossens (2012) in the limit of small twists showed that the magnetic twist changes the polarization of the transverse motions of standing kink oscillation along the flux tube but does not affect its frequency. Ruderman (2015) investigated the effect of a continues twisted magnetic field on the propagating kink modes in the thin tube approximation. He showed that there are two propagating kink waves with the same longitudinal wave numbers but opposite propagation directions, which have different frequencies. Ruderman (2015) called these waves ``accelerated'' and ``decelerated'' kink waves which have larger and smaller frequencies with respect to the well known kink frequency, respectively.

Karami \& Bahari (2010) studied the effect of twisted magnetic
field on the resonant absorption of MHD waves due to the radial density structuring in
coronal loops. They showed that by increasing the twist parameter, the frequency, the damping rate and their ratio for both the kink and fluting modes increase. Also the magnetic twist causes the ratio of fundamental period to first overtone one for
kink and fluting modes to be smaller than 2.

The main goal of the present work is to study the resonant
absorption of kink MHD waves by magnetic twist to explain the rapid damping of oscillating coronal loops and departure of the
period ratios $P_1/P_2$ and $P_1/P_3$ from their canonical values
reported by observations. To this aim, in section \ref{sec2} we
introduce the coronal loop model and find the solutions of the
equations of motion. In section \ref{sec3}, we investigate a number of possible instabilities that may arise in our model. In section \ref{sec4}, we use the appropriate
connection formulae to obtain the dispersion relation. In section
\ref{sec5}, we solve the dispersion
relation, analytically. Section \ref{sec6} gives the summary and conclusions.

\section{Model and equations of motion}\label{sec2}

As a simplified model for a coronal loop, we
consider a straight cylindrical flux tube with
length $L$ and radius $R$. The background density profile is
assumed to be
\begin{eqnarray}\label{rho}
\rho(r)=\left\{\begin{array}{lll}
    \rho_{{\rm i}},&0<r\leq R,&\\
    \rho_{{\rm e}},&r>R,&\\
      \end{array}\right.
\end{eqnarray}
where $\rho_{\rm i}$ and $\rho_{\rm e}$ are the constant densities of the interior and exterior regions of the flux tube, respectively. We define the density ratio $\zeta\equiv\rho_{\rm i}/\rho_{\rm e}$ in the rest of the paper.

We further assume that the background magnetic field to have a small twist in a thin layer and to be constant and aligned with the loop axis everywhere else, i.e.
\begin{eqnarray}\label{B}
\mathbf{B}=\left\{\begin{array}{lll}
    B_{{\rm 0}}\hat{z},&0<r<a,&\\
    B_{{ \varphi}}(r)\hat{\varphi}+B_0\hat{z},&a\leq r\leq R,&\\
    B_{{\rm 0}}\hat{z},&r>R.&\\
      \end{array}\right.
\end{eqnarray}
Here, we set a parabolic profile for $B_\varphi$ as
follows
\begin{eqnarray}
B_{\varphi}(r)=Ar(r-a),\label{Bphi}
\end{eqnarray}
where $A$ is a constant. Note that the jump in
the value of $B_{{ \varphi}}(r)$ across $r=R$ implies a delta-function current sheet there.

The magnetohydrostatic equilibrium equation
takes the form
\begin{eqnarray}
\frac{{\rm d} }{{\rm d}
r}\left(P+\frac{B_{\varphi}^2+B_z^2}{2\mu_0}\right)=-\frac{B_{\varphi}^2}{\mu_0 r},\label{mse}
\end{eqnarray}
where $\mu_0$ is the magnetic permeability of free space. Then, by integrating Eq. (\ref{mse}) and using the continuity of the total (magnetic plus gas) pressure at $r=a$ and $r=R$, we can find the gas pressure as follows
\begin{eqnarray}
P(r)=\left\{\begin{array}{lll}
            \frac{B_{\rm 0}^2}{2\mu_0}\beta,&0<r< a,\\
            \frac{B_{\rm 0}^2}{2\mu_0}\beta-\frac{A^2}{12\mu_0}(r-a)\\
            \times(9r^3-11ar^2+a^2r+a^3),&a\leq r\leq R,\\
            \frac{B_{\rm 0}^2}{2\mu_0}\beta-\\
            \frac{A^2}{12\mu_0}(3R^4-8aR^3+6a^2R^2-a^4),&r> R,\\
      \end{array}\right.\label{Pg}
\end{eqnarray}
where parameter $\beta\equiv \frac{P}{B_{\rm 0}^2/(2\mu_0)}$is the ratio of the plasma pressure to the magnetic field pressure, inside the loop.

\begin{figure}
\centering
    \includegraphics[width=0.48\textwidth]{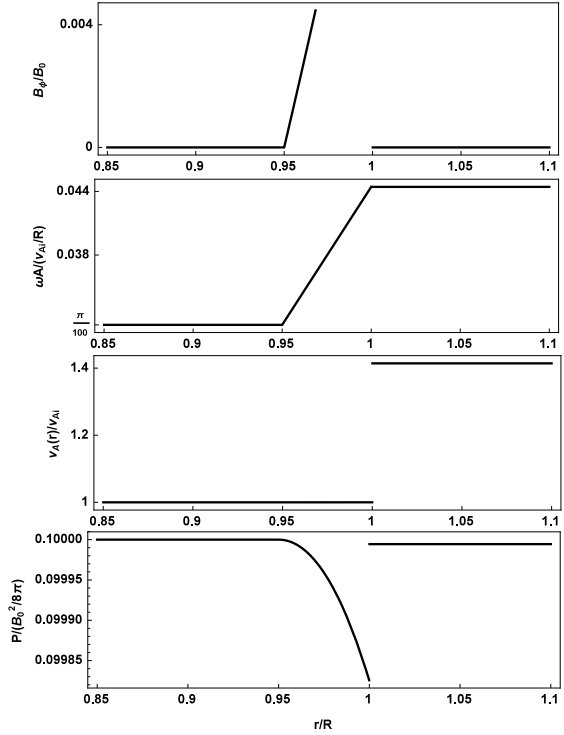}
    \caption[] {Azimuthal component of the background magnetic field, Alfv\'{e}n frequency of the fundamental ($l=1$) kink ($m=1$) mode, background Alfv\'{e}n speed and background plasma pressure versus the fractional radius $r/R$. The loop parameters are:
      $L=10^{5}$ km, $R/L=0.01$, $a/R=0.95$, $\zeta=2$, $\rho_{\rm i}=2\times10^{-14}$ g $\rm{cm^{-3}}$, $\beta=0.1$ and
      $B_0=100$ G. Here $\alpha=\alpha_{\rm c}=0.013$ and $v_{A_{\rm i}}=B_0/\sqrt{\mu_0\rho_{i}}=2000$ km s$^{-1}$.}
    \label{background}
 \end{figure}

The linearized ideal MHD equations for incompressible plasma are given by
\begin{eqnarray}
\frac{\partial\delta{\mathbf v}}{\partial
t}=-\frac{\nabla\delta P}{\rho}+\frac{1}{\mu_0\rho}\{(\nabla\times\delta{\mathbf
B})\times{\mathbf B} +(\nabla\times{\mathbf B})\times\delta{\mathbf
B}\},\label{mhd1}
\end{eqnarray}
\begin{eqnarray}
\frac{\partial\delta{\mathbf B}}{\partial
t}=\nabla\times(\delta{\mathbf v}\times{\mathbf B}),\label{mhd2}
\end{eqnarray}
\begin{eqnarray}
\nabla \cdot \delta \mathbf v=0,\label{mhd3}
\end{eqnarray}
where $\delta\bf{v}$, $\delta\bf{B}$ and $\delta P$ are the Eulerian
perturbations of velocity, magnetic fields and plasma pressure; $\bf{B}$ and $\rho$ are the background magnetic filed and the mass density, respectively. Also $t$-, $\phi$-
and $z$-dependency of the perturbations are supposed to be of the
form $\exp[i(m\phi+k_z z-\omega t)]$ where $k_z=\frac{l\pi}{L}$ is
the longitudinal wave number. Here $m$ and $l$ are the azimuthal and
longitudinal mode numbers, respectively, and $\omega$ is the mode
frequency. So the perturbations are of the form
\begin{eqnarray}
\delta {\mathbf v}(r,\phi,z,t)=\delta {\mathbf v}(r)\exp[i(m\phi+k_z z-\omega t)],\nonumber \\
\delta {\mathbf B}(r,\phi,z,t)=\delta {\mathbf B}(r)\exp[i(m\phi+k_z z-\omega t)],\nonumber\\
\delta {\rm P}(r,\phi,z,t)=\delta {\rm P}(r)\exp[i(m\phi+k_z z-\omega t)].\label{pert}
\end{eqnarray}
Following Bennett et al. (1999), substituting the perturbations (\ref{pert}) into Eqs.
(\ref{mhd1}) and (\ref{mhd2}) and doing some algebra we get
\begin{subequations}
    \begin{empheq}{align}
       & \frac{\rm d^2}{{\rm d} r^2}\delta p+\left\{\frac{C_3}{r D}\frac{\rm d}{{\rm d}r}\left(\frac{r D}{C_3} \right) \right\}\frac{\rm d}{{\rm d}r}\delta p\nonumber\\
       &+\left \{\frac{C_3}{r D}\frac{\rm d}{{\rm d}r}\left(\frac{r C_1}{C_3} \right)+\frac{1}{D^2}(C_2C_3-C_1^2) \right\}\delta p=0,\label{p}\\
       & \xi_r=\frac{D}{C_3}\frac{\rm d}{{\rm d}r}\delta p+\frac{C_1}{C_3}\delta p,\label{xi}
    \end{empheq}
\end{subequations}
where $\delta p=\delta P+\mathbf B\cdot\delta\mathbf B / \mu_0$ and
$\xi_r=-\delta v_r/i\omega$ are the Eulerian perturbation of
total pressure and the Lagrangian displacement in the radial
direction, respectively and
\begin{eqnarray}\label{DCs}
    \begin{aligned}
       & D=\rho(\omega^2-\omega_{\rm A}^2),\\
       & C_1=-\frac{2 m B_\varphi}{\mu_0 r^2}\left(\frac{m}{r}B_\varphi+k_zB_z \right),\\
       & C_2=-\left(\frac{m^2}{r^2}+k_z^2 \right),\\
       & C_3=D^2+D\frac{2B_\varphi}{\mu_0}\frac{\rm d}{\rm{d}r}\left(\frac{B_\varphi}{r}\right)-\frac{4B_\varphi^2}{\mu_0 r^2}\rho  \omega_A^2.
    \end{aligned}
\end{eqnarray}
Here, the Alfv\'{e}n frequency, $\omega_A$, is defined as
\begin{eqnarray}\label{omegaAA}
\omega_A(r)\equiv\frac{1}{\sqrt{\mu_0\rho}}\left(\frac{m}{r}B_\varphi+k_z B_z \right).
\end{eqnarray}
Putting Eqs. (\ref{rho})-(\ref{Bphi}) into (\ref{omegaAA})
gives the profile of Alfv\'{e}n frequency as follows
\begin{eqnarray}\label{omegaA2}
\omega_A(r)=\left\{\begin{array}{lll}
    \frac{B_{\rm 0}k_z}{\sqrt{\mu_0\rho_{\rm i}}},&0<r< a,&\\
    \frac{1}{\sqrt{\mu_0\rho_{\rm i}}}\Big(m A (r-a)+k_zB_0\Big),&a\leq r\leq R,&\\
    \frac{B_{\rm 0}k_z}{\sqrt{\mu_0\rho_{\rm e}}},&r> R.&\\
      \end{array}\right.
\end{eqnarray}
Here we use the twist parameter defined as $\alpha\equiv\frac{B_\varphi(R)}{B_0}$. For a special value of $\alpha=\alpha_{\rm c}$, the Alfv\'{e}n frequency is continues at the tube boundary ($r=R$). If $\alpha\neq\alpha_{\rm c}$ there would be a gap in the Alfv\'{e}n frequency profile across the boundary. It is straightforward from Eq. (\ref{omegaA2}) to show that
\begin{equation}\label{alphamin}
    \alpha_{\rm c}=\frac{R k_z}{m}\left(\sqrt{\zeta}-1\right).
\end{equation}
Figure \ref{background} shows the azimuthal component of the background magnetic field, the Alfv\'{e}n frequency of the fundamental ($l=1$) kink ($m=1$) mode, the background Alfv\'{e}n speed, $v_A(r)=B/\sqrt{\mu_0\rho}$, and the background plasma pressure for the twist parameter $\alpha=\alpha_{\rm c}=0.013$. Note that although for $\alpha>\alpha_{\rm c}$, there is a sudden drop in the Alfv\'{e}n frequency $\omega_A(r)$ at the tube boundary, the interior Alfv\'{e}n speed $v_A(r)$ in the present model is still smaller than the exterior one which is likely to occur under coronal conditions.

From Eqs. (\ref{p})-(\ref{omegaAA}), it is clear that when $\omega^2=\omega_{A}^2$, i.e. $D=0$, the equations of motion in the
inhomogeneous layer ($a<r<R$) become singular in the presence of
magnetic twist. Therefore, resonant absorption can occur not
only by the radial density inhomogeneity (like previous works,
see e.g. Ruderman \& Roberts 2002; Karami et al. 2009; Karami \& Bahari 2010; Ruderman 2011; Ruderman \& Terradas 2013; Soler \& Terradas 2015) but also by the magnetic twist (present
work).

In the untwisted regions, $0<r<a$ and $r>R$, Eqs. (\ref{p}) and (\ref{xi}) are reduced to the following equations
\begin{subequations}
    \begin{empheq}{align}
        &\frac{{\rm d}^2\delta p}{{\rm d}r^2}+\frac{1}{r}\frac{{\rm d}\delta p}{{\rm d} r}-\left(k^2_{z}+\frac{m^2}{r^2}\right)\delta p=0,\label{peq} \\
        &\xi_r=\frac{1}{\rho(\omega^2-\omega_A^2)}\frac{{\rm d} \delta p}{{\rm d} r}.\label{xireq}
    \end{empheq}
\end{subequations}
Solutions for equations
(\ref{peq}) and (\ref{xireq}) in the interior ($0<r<a$) and exterior ($r>R$) regions are
obtained as
\begin{subequations}\label{inoutsol}
    \begin{empheq}{align}
        &\delta p(r)=\left\{\begin{array}{lll}
        A_{\rm i} I_{\rm m}(k_z r),&0<r<a, &\\
        A_{\rm e} K_{\rm m}(k_z r),&r>R,&\\
         \end{array}\right.\label{inoutsol1}
         \\
        &\xi_r(r)=\left\{\begin{array}{lll}
        A_{\rm i}\frac{k_z}{\rho_{\rm i}(\omega^2-\omega_{\rm A_i}^2)} I'_{\rm m}(k_z r),&0<r<a,&\\
        A_{\rm e}\frac{k_z}{\rho_{\rm e}(\omega^2-\omega_{\rm A_e}^2)} K'_{\rm m}(k_z r),&r>R.&\\
        \end{array}\right.\label{inoutsol2}
    \end{empheq}
\end{subequations}
Here $I_{\rm m}$ and $K_{\rm m}$ are the modified Bessel functions of the first and second kind, respectively. Also ``$'$''
on $I_{\rm m}$ and $K_{\rm m}$ represents a derivative with respect
to their arguments. The constants $A_{\rm i}$ and
$A_{\rm e}$ are determined by the appropriate boundary
conditions.

\section{Stability constraints on the model}\label{sec3}
In what follows, we are interested in investigating a number of possible instabilities that may arise in our model.

\subsection{Kink instability}
If the magnetic twist value in the loop, which is defined as $\phi_{\rm twist}\equiv\frac{L}{R}\frac{B_{\varphi}}{B_z}=2\pi N_{\rm twist}$ exceeds a critical value $\phi_{\rm c}$, then the loop becomes kink unstable (see e.g. Shafranov 1957; Kruskal et al. 1958). Here, $N_{\rm twist}$ is the number of twist turns in the loop. For force-free magnetic fields of
uniform twist, a critical twist of $\phi_{\rm twist}\gtrsim 3.3\pi$ (or 1.65 turns) was found to lead to
kink instability (Hood \& Priest 1979), while the critical value ranges between $2\pi$ and $6\pi$ for other types of
magnetic fields (see e.g. Aschwanden 2005; Priest 2014). Numerical MHD simulations of an increasingly
twisted loop system demonstrated linear instability of the ideal MHD kink
mode for twist angles in excess of $\thickapprox 4.8\pi$ (or 2.4 turns) (Miki\'c et al. 1990). Baty \& Heyvaerts (1996) examined the kink instability of a radially localized twist profile and obtained $\phi_{\rm c}=5\pi$.
In our model to avoid the kink instability, following Hood \& Priest (1979) we consider $\phi_{\rm c}=3.3\pi$ (or 1.65 turns). This yields an upper limit for the twist parameter as $\alpha_{\rm max}=(R/L)\phi_{\rm c}$.

\subsection{Kelvin-Helmholtz instability}
The Kelvin-Helmholtz instability (KHI) can occur during the kink oscillations of coronal loops. The kink wave, whose initial energy is mostly
transverse, converts into azimuthal waves locally resembling
torsional Alfv\'{e}n waves, which are finely localized around the
tube’s boundary layer (Verth et al. 2010; Arregui et al. 2011).
Azimuthal motions are thus amplified and introduce velocity
shear, making them prone to be unstable to the KHI (Heyvaerts \& Priest 1983; Soler
et al. 2010). Terradas et al. (2008) using numerical simulations showed that azimuthal shear
motions generated at the loop boundary during kink oscillations can give rise to a KHI, but this phenomenon has not been
observed to date. The KHI can also extract the energy from the resonant
layer and convert it into heat through viscous and ohmic
dissipations at the generated vortices and current sheets (Antolin et al. 2015). Soler et al. (2010) pointed out that the presence of a small azimuthal component of the magnetic field can suppress the KHI in a stable coronal loop. They showed that the required twist is small enough to prevent the development of the pinch instability. A weak
twist of magnetic field lines is very likely and realistic in the
context of coronal loops.

\subsection{Resistive kink instability}\label{reskink}
\begin{figure}
\centering
    \includegraphics[width=0.48\textwidth]{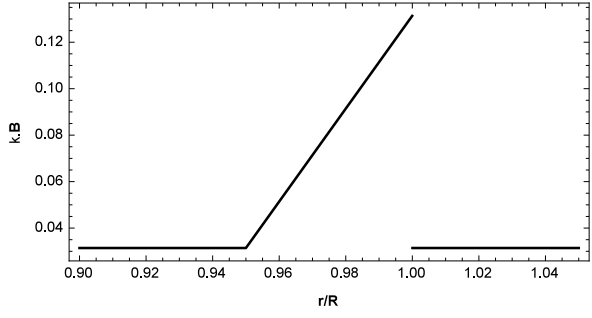}
    \caption[] {Variation of $\mathbf{k}\cdot \mathbf{B}$ for the fundamental ($l=1$) kink ($m=1$) mode versus the fractional radius $r/R$. Here $\alpha=0.1$ and the other auxiliary parameters as in Fig. \ref{background}.}
    \label{kb}
 \end{figure}

The effect of resistive diffusion on the ideal kink mode yields a so-called resistive kink instability which is a reconnecting process. The resistive kink becomes important when the loop is twisted too much (see e.g. Biskamp 2000; Wesson 2004; Priest 2014). From Eq. (\ref{mhd2}), it is possible to have
\begin{equation}\label{rk1}
   \nabla\times(\delta{\mathbf v}\times{\mathbf B})=(\mathbf{B}\cdot\nabla)\delta\mathbf{v}-\mathbf{B}(\nabla\cdot\delta \mathbf{v})=0,
\end{equation}
 at a specific location $r=r_{\rm s}$. In this case, the diffusion term $\eta\nabla^2\delta \mathbf B$ neglected from the right hand side of Eq. (\ref{mhd2}) becomes important and consequently the field lines diffuse through the plasma and reconnect. Here $\eta$ is the magnetic diffusivity. For an incompressible plasma, inserting Eq. (\ref{mhd3}) into  (\ref{rk1}) gives rise to $\mathbf{k}\cdot\mathbf{B}\big|_{r=r_{\rm s}}=0$, where $\mathbf{k}$ and $\mathbf{B}$ are the wave vector and the background magnetic field, respectively. Using Eq. (\ref{B}), the necessary condition for the resistive kink instability takes the form
\begin{equation}\label{tearing}
 \mathbf k\cdot \mathbf{B}\big|_{r=r_{\rm s}}=\frac{B_{\varphi}(r_{\rm s})}{r_{\rm s}}+k_z B_z(r_{\rm s})=0.
\end{equation}
In Fig. \ref{kb}, we plot $\mathbf{k}\cdot\mathbf{B}$ for the fundamental ($l=1$) kink ($m=1$) mode. Figure illustrates that the term $\mathbf{k}\cdot\mathbf{B}$ cannot be zero and consequently the resistive kink instability
cannot occur in our model.

\subsection {Tearing mode instability}
The tearing mode instability can occur in a thin current sheet when the driving force of the inflow exceeds the opposing Lorentz force (see e.g. Furth et al. 1963; Goldstone \& Rutherford 1995; Magara \& Shibata 1999; Tenerani et al. 2015). As a consequence of non-zero resistivity, the magnetic field lines tear and reconnect in the current sheet. According to Furth et al. (1973), the smallest growth time of the tearing mode in the cylindrical flux tube is given by about
\begin{equation}\label{teartime}
    \tau_t\sim\tau_A^{2/5}\tau_d^{3/5},
\end{equation}
where $\tau_{\rm A}=l_{\rm s}/v_{A_{\rm i}}$ and $\tau_d=l_{\rm s}^2/\eta$ are the Alfv\'{e}n and magnetic diffusion time scales, respectively. Here $l_{\rm s}$ is a half-thickness of the current sheet and $v_{A_{\rm i}}=B_0/\sqrt{\mu_0\rho_{i}}$ is the interior Alfv\'{e}n
speed. In our model, due to having a rotational discontinuity of the background magnetic field at the tube boundary, we have a delta-function current sheet. Therefore, from Eq. (\ref{teartime}) when $l_{\rm s}\rightarrow 0$ the growth time of the tearing process goes to zero. It should be noted that the delta-function current sheet in our model is used to approximate a finite width current layer. In reality, the thickness of such current sheets in the solar corona takes place in the range of macroscopic values (10 km, for example), see e.g. Magara \& Shibata (1999). For the present model, if we take $T=5\times 10^6$ K, $\eta=10^9 T^{-3/2}=0.09$ m$^2$ s$^{-1}$ (Priest 2014), $v_{A_{\rm i}}=2000$ km s$^{-1}$ then for $2l_0\simeq(1-10)$ km (Magara \& Shibata 1999) we estimate the tearing growth time as  $\tau_t/P_{\rm kink}\sim (3-123)$. Here  $P_{\rm kink}=2\pi/\omega_{\rm kink}\simeq 87$ s is the period of the fundamental kink mode oscillation (see Eq. \ref{omegakink}). Therefore, by choosing an appropriate thickness for the current sheet, the tearing mode instability can be avoided in our model during the resonant absorption of the kink oscillations.

It is worth noting that considering a continuous background magnetic field like the profile used in the work of Hood et al. (2016), allows us to have a twist in the tube without having the delta-function current sheet. This will make it easier to avoid the growth of the tearing instability. However, for a certain density distribution (like the one we use in this paper) the model of Hood et al. (2016) does not yield a monotonic function of the background Alfv\'{e}n frequency. As a result it may give rise to two resonant layers. In this case, in order to use the connection formulae (see section 4) at the resonance layers, we need to solve the MHD equations in the twisted regions around the resonance points. However, this is beyond the scope of this paper and we leave it for future work.

\section{Connection formulae and dispersion relation }\label{sec4}

\begin{figure}
\centering
    \includegraphics[width=0.48\textwidth]{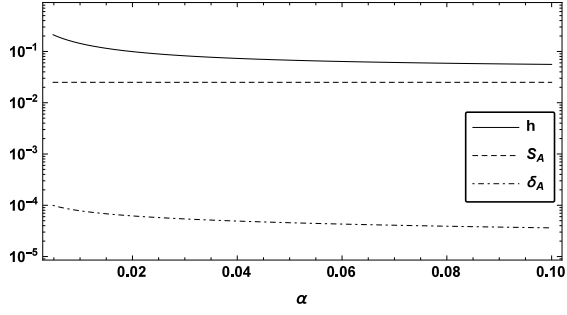}
    \caption[] {Variations of $h$, $S_{\rm A}=0.025$ and thickness of the resonance layer $\delta_{\rm A}$ versus the twist parameter $\alpha$ for the fundamental ($l=1$) kink ($m=1$) modes. Here $h$, $\delta_A$ and $S_A$ are in units of the loop radius $R=1000$ km.  Auxiliary parameters as in Fig. \ref{background}.}
    \label{width}
 \end{figure}

Here, we do not solve Eq. (\ref{p}) in the inhomogeneous
layer ($a\leq r\leq R$), where the singularity occurs due to the existence of
the magnetic twist. Instead, the solutions inside and outside of the
tube can be related to each other by the connection formulae
introduced by Sakurai et al. (1991a). To check the validity of the connection formulae, the radius $S_A$ of the region around the resonance point that connects the solutions of the perturbations of the interior and exterior of the flux tube, must obey the following condition (see Goossens et al. 2011)
\begin{equation}\label{con1}
   S_A\ll h \equiv \left |\frac{\frac{\rm d}{\rm dr}~\omega_A^2(r)}{\frac{\rm d^2}{\rm dr^2}~\omega_A^2(r)}\right|_{r=r_{\rm A}}.
\end{equation}
In our work, we set $S_A=\frac{R-a}{2}$. In Fig. \ref{width}, we plot $S_A$ and $h$ versus the twist parameter $\alpha=\frac{B_\varphi(R)}{B_0}$. Figure clears that for the range of magnetic twist considered in this study, the condition (\ref{con1}) holds.  Therefore, in the thin boundary approximation we
can use the connection formulae around the inhomogeneous layer. According to Sakurai et al.
(1991a) the jumps across the boundary (resonance layer) for $\delta
p$ and $\xi_r$ are
\begin{subequations}
    \begin{empheq}{align}
        &[\delta p]=-\frac{i\pi}{|\Delta|}\frac{2B_\varphi(r_A)B_z(r_A)f_B(r_A)}{\mu_0\rho_{\rm i}r_A B^2(r_A)}C_A(r_A),\label{jump1}\\
        &[\xi_r]=-\frac{i\pi}{|\Delta|}\frac{g_B(r_A)}{\rho_{\rm i}
        B^2(r_A)}C_A(r_A),\label{jump2}
    \end{empheq}
\end{subequations}
where
\begin{eqnarray}
    \begin{aligned}\label{fgDelta}
    &C_A=g_B \delta p(r)-\frac{2f_B B_\varphi B_z}{\mu_0 r_A} \xi_r(r),\\
    &f_B=\frac{m}{r}B_\varphi+k_z B_z,\\
    &g_B=\frac{m}{r}B_z-k_z B_\varphi,\\
    &\Delta=-\frac{\rm d}{{\rm d}r}\omega_A^2(r),
    \end{aligned}
\end{eqnarray}
and $r_{A}$ is the location of the resonance point.
Notice that the ideal MHD solutions for the untwisted regions, i.e. inside ($0<r<a$) and outside ($r>R$) the flux tube, are valid in the jump conditions (\ref{jump1}) and (\ref{jump2}) only if the following condition is established (see Stenuit et al. 1998; Goossens et al. 2011)
\begin{equation}\label{deltaA}
   \delta_A\leq S_A=\frac{R-a}{2},
\end{equation}
 where $\delta_A=\big|\frac{\omega}{|\Delta|}(\nu+\eta)\big|^{1/3}=(v_{A_{\rm i}}R)^{1/3}\big|\frac{\omega}{|\Delta|}(\frac{1}{\mathcal{R}_v}+\frac{1}{\mathcal{R}_m})\big|^{1/3}$ is a measure of the thickness of the resonance layer (see Sakurai et al. 1991a). Here $\nu=\frac{v_{A_{\rm i}}R}{\mathcal{R}_v}$, $\eta=\frac{v_{A_{\rm i}}R}{\mathcal{R}_m}$ and $\mathcal{R}_v$ are the kinematic viscosity, magnetic diffusivity and viscus Reynolds number, respectively. The classical values of viscus and resistive Reynolds numbers of the solar corona are about $10^{14}$ and $10^{13}$, respectively (see Colub \& Pasachoff 1997). In Fig. \ref{width}, we also plot $\delta_A$ for the aforementioned values of the Reynolds and Lundquist numbers. Figure clears that the condition (\ref{deltaA}) is respected.

Substituting the ideal solutions (\ref{inoutsol1}) and (\ref{inoutsol2})
in the jump conditions (\ref{jump1}) and (\ref{jump2}) and
eliminating $A_{\rm i}$ and $A_{\rm e}$, one can find the
dispersion relation as
\begin{equation}\label{disp}
    d_0(\tilde\omega)+d_1(\tilde\omega)=0,
\end{equation}
where
\begin{eqnarray}\label{d0d1}
    \begin{aligned}
    d_0(\tilde\omega)=\frac{k_z}{\rho_{\rm e}(\tilde\omega^2-\omega_{A_{\rm e}}^2)}\frac{K'_{\rm m}(k_z R)}{K_{\rm m}(k_z R)}\\
    ~~~~~~~~-\frac{k_z}{\rho_{\rm i}(\tilde\omega^2-\omega_{A_{\rm i}}^2)}\frac{I'_{\rm m}(k_z a)}{I_{\rm m}(k_z a)}, \end{aligned}
\end{eqnarray}
and
\begin{eqnarray}\label{d0d1}
    \begin{aligned}
    &d_1(\tilde\omega)=\\
    &\frac{i\pi}{|\Delta|} \left[\frac{g_B}{\rho_{\rm i}B_0^2}-\frac{k_z}{\rho_{\rm e}(\tilde\omega^2-\omega_{A_{\rm e}}^2)}\frac{2 f_B B_{\varphi}}{\mu_0\rho_{\rm i} r_{\rm A} B_0}\frac{K'_{\rm m}(k_z R)}{K_{\rm m}(k_z R)} \right]\\
    &\times \left[g_B\frac{I_m(k_z r_{\rm A})}{I_m(k_z a)}-\frac{k_z}{\rho_{\rm i}(\tilde\omega^2-\omega_{A_{\rm i}}^2)}\frac{2f_B B_{\varphi} B_0}{\mu_0 r_{\rm A}}\frac{I'_{\rm m}(k_z r_{\rm A})}{I_{\rm m}(k_z a)}\right].
    \end{aligned}
\end{eqnarray}
Here $\tilde\omega=\omega-i\gamma$, in which $\omega$ and
$\gamma$ are the mode frequency and the corresponding damping rate,
respectively.

Using Eq. (\ref{omegaA2}) in the last relation of Eq. (\ref{fgDelta}) we get
\begin{equation}\label{Delta}
    \Delta=-\frac{B_0^2}{\mu_0\rho_{\rm i}}\frac{2m\alpha}{R(R-a)}\left(m\alpha\frac{r-a}{R(R-a)}+k\right),
\end{equation}
which shows that in the limit of $a\rightarrow R$, i.e. when the twisted annulus region is removed, then $\Delta$ goes to infinity. As a result, in Eq. (\ref{d0d1}), $d_1=0$. In this case there is no resonant absorption and the dispersion relation (\ref{disp}) is the same as Eq. (25) in Bennett et al. (1999).

In the limit of thin tube approximation, using the first order asymptotic expansions for $I_{\rm m}(x)$ and $K_{\rm m}(y)$ one can get
\begin{eqnarray}\label{TT}
    \begin{aligned}
    &\frac{I_{\rm m}(k_z r_{\rm A})}{I_{\rm m}(k_z a)}\simeq \left(\frac{r_{\rm A}}{a} \right)^m,\\
    &\frac{I'_{\rm m}(k_z r_{\rm A})}{I_{\rm m}(k_z a)}\simeq\frac{m}{k_z r_{\rm A}}\left(\frac{r_{\rm A}}{a} \right)^m ,\\
    & \frac{I'_{\rm m}(k_z a)}{I_{\rm m}(k_z a)}\simeq\frac{m}{k_z a},\\
    &\frac{K'_{\rm m}(k_z R)}{K_{\rm m}(k_z R)}\simeq-\frac{m}{k_z R}.
    \end{aligned}
\end{eqnarray}
Using the above approximations, the dispersion relation (\ref{disp}) reduces to
\begin{eqnarray}\label{disptt}
    \begin{aligned}
     &\frac{\rho_{\rm i}(\tilde\omega^2-\omega_{A_{\rm i}}^2)}{R}+\frac{\rho_{\rm e}(\tilde\omega^2-\omega_{A_{\rm e}}^2)}{a}\\
     &~~~~~~~~-\frac{i\pi}{|\Delta|}\left(\frac{r_{\rm A}}{a}\right)^m\left(\frac{g_{\rm B}}{B_0^2}\rho_{\rm e}(\tilde\omega^2-\omega_{A_{\rm e}}^2)
    +\frac{2mf_{B}B_{\varphi}}{\mu_0 r_{\rm A} B_0 R} \right)\\
    &~~~~~~~~~~~~~~~~~\times\left(\frac{g_{\rm B}}{m}(\tilde\omega^2-\omega_{A_{\rm i}}^2)-\frac{2f_{B}B_{\varphi}B_0}{\mu_0 \rho_{\rm i} r_{\rm A}^2}
    \right)=0.
    \end{aligned}
\end{eqnarray}
Notice that we look for the frequencies in the range of
$\omega_{A_{\rm i}}< \omega < \omega_{A_{\rm e}}$, in which the
resonant absorption occurs. Here $\omega_{A_{\rm
i}}=k_{z}\frac{B_{\rm 0}}{\sqrt{\mu_{\rm 0}\rho_{\rm i}}}$ and
$\omega_{A_{\rm e}}=k_{z}\frac{B_{\rm 0}}{\sqrt{\mu_{\rm 0}\rho_{\rm
e}}}$ are the interior and exterior Alfv\'{e}n frequencies,
respectively.
In the next section, we solve Eq. (\ref{disptt}) analytically to obtain
the frequency and damping rate of the kink MHD modes.
\section{Analytical results}\label{sec5}

Here, we are interested in studying the effect of the twist
parameter $\alpha$ on the frequencies, $\omega$, and the damping rates,
$\gamma$, of the kink ($m=1$) MHD modes in a resonantly damped
coronal loop. To this aim, we need to solve Eq.
(\ref{disptt}). First, we use the dimensionless
quantities $\bar{r}=r/R$, $\bar{L}=L/R$, $\bar{B}=B/B_0$
and $\bar{\tilde{\omega}}=\tilde{\omega}/(v_{A_{\rm i}}/R)$. Thus, Eq. (\ref{disptt}) can be recast in
dimensionless form as
\begin{eqnarray}\label{dispttdim}
\begin{aligned}
 &\zeta(\tilde\omega^2-\omega_{A_{\rm i}}^2)+\frac{1}{q}(\tilde\omega^2-\omega_{A_{\rm e}}^2)\\
 &~~~~~~~~-\frac{i\pi}{|\Delta|}\left(\frac{r_{\rm A}}{a}\right)^m\left(g_{\rm B}(\tilde\omega^2-\omega_{A_{\rm e}}^2)+\frac{2\zeta mf_{B}B_\varphi}{r_{\rm A}} \right)\\
 &~~~~~~~~~~~~~~~~\times\left(\frac{g_{\rm B}}{m}(\tilde\omega^2-\omega_{A_{\rm i}}^2)-\frac{2 f_{B}B_\varphi}{r_{\rm A}^2} \right)=0,
 \end{aligned}
\end{eqnarray}
where we have dropped the bars for simplicity. Here $q\equiv a/R$. Equation (\ref{dispttdim}) yields a quadratic equation for $\tilde\omega^2$ as follows
\begin{eqnarray}\label{quadratic}
   c_1\tilde\omega^4+c_2\tilde\omega^2+ c_3=0,
\end{eqnarray}
where
\begin{eqnarray}\label{quadcof}
    \begin{aligned}
   &c_1=-\frac{i \pi}{|\Delta|}\left(\frac{r_{\rm A}}{q} \right)^m\frac{g_B^2}{m},\\
   &c_2=\left(\zeta+\frac{1}{q}\right)+\frac{i \pi}{|\Delta|}\left(\frac{r_{\rm A}}{q} \right)^m\\
   &\times\left[ \frac{g_B^2}{m}\left(\omega_{A_{\rm i}}^2+\omega_{A_{\rm e}}^2\right)+\frac{2 f_B g_B B_\varphi}{r_{\rm A}^2}\left(1-\zeta r_{\rm A}\right)\right],\\
   &c_3=-\zeta\omega_{A_{\rm i}}^2\left( 1+\frac{1}{q}\right)-\frac{i \pi}{|\Delta|}\left(\frac{r_{\rm A}}{q} \right)^m\zeta\\
   & \times\left[\frac{g_B^2}{m}\omega_{A_{\rm i}}^4-\frac{4 m f_B^2 B_\varphi^2}{r_{\rm A}^3}+\frac{2 f_B g_B B_\varphi\omega_{A_{\rm i}}^2}{r_{\rm A}^2}(1-r_{\rm A}) \right].
   \end{aligned}
\end{eqnarray}
It is straightforward to find the solutions of Eq. (\ref{quadratic}), $\tilde\omega=\omega-i\gamma$, as follows
\begin{eqnarray}\label{sol}
    \begin{aligned}
        &\omega_\pm=\frac{1}{\sqrt{\frak{R}_{\pm}^2+\frak{I}_{\pm}^2}}\left(\sqrt{\frak{\chi_{\pm}}^2+\Theta_{\pm}^2}+\frak{\chi_{\pm}}\right)^{1/2},\\
        &\gamma_\pm=\frac{1}{\sqrt{\frak{R}_{\pm}^2+\frak{I}_{\pm}^2}}\left(\sqrt{\frak{\chi_{\pm}}^2+\Theta_{\pm}^2}-\frak{\chi_{\pm}}\right)^{1/2},
    \end{aligned}
\end{eqnarray}
where
\begin{eqnarray}\label{RI}
    \begin{aligned}
        &\chi_{\pm}=c_{3_{\rm R}}\frak{R}_{\pm}+c_{3_{\rm I}}\frak{I}_{\pm},\\
        &\Theta_{\pm}=c_{3_{\rm I}}\frak{R}_{\pm}-c_{3_{\rm R}}\frak{I}_{\pm},\\
        & \frak{R_{\pm}}=-c_{2_{\rm R}}\pm\left(\frac{\sqrt{c_{\rm 4}^2+c_{\rm 5}^2}+c_{\rm 4}}{2}\right)^{1/2},\\
        & \frak{I_{\pm}}=-c_{2_{\rm I}}\pm\left(\frac{\sqrt{c_{\rm 4}^2+c_{\rm 5}^2}-c_{\rm 4}}{2}\right)^{1/2},
    \end{aligned}
\end{eqnarray}
with
\begin{eqnarray}\label{crci}
    \begin{aligned}
        & c_4=c_{2_{\rm R}}^2-c_{2_{\rm I}}^2+4c_{1_{\rm I}}c_{3_{\rm I}},\\
        &c_5=2c_{2_{\rm R}}c_{2_{\rm I}}-4c_{1_{\rm I}}c_{3_{\rm R}}.
    \end{aligned}
\end{eqnarray}
Here, the subscripts $R$ and $I$ denote the real and imaginary parts of a given complex quantity.

For a typical coronal loop, we take $L=10^5$
km, $R/L=0.005,0.01$, $\zeta=\rho_{\rm i}/\rho_{\rm e}=2,4$, $\rho_{\rm
i}=2\times10^{-14}$  g $\rm {cm^{-3}}$ and $B_0=100$ G. So, one finds $v_{A_{\rm i}}=B_0/\sqrt{\mu_0\rho_{\rm i}}=2000$ km $\rm{s^{-1}}$ in this loop. We
assume that the magnetic twist takes place in a thin layer of
thickness $R-a=0.05R$. So we have $q=0.95$. In the thin boundary approximation, we can set the location of the resonance point at the tube surface, i.e. $r_{\rm A}\simeq R$.

Using Eq. (\ref{sol}) we obtain two roots of $\omega_{\pm}$ which their values take place in the range of $\omega_{A_{\rm i}}<\omega_{-}<\omega_{A_{\rm e}}$ and $\omega_{+}>\omega_{A_{\rm e}}$.
Therefore, $\omega_{-}$ is our physical root and $\omega_{+}$ should be ruled out, because it does not give rise to the resonant absorption.
As we have already mentioned, in the limit of $q\rightarrow 1$ (i.e. $a\rightarrow R$), we have $\Delta\rightarrow\infty$. In this case, from Eq. (\ref{quadcof}) we obtain $c_1=c_{2_{\rm I}}=c_{3_{\rm I}}=0$. Hence, in the absence of twisted annulus region, form Eqs. (\ref{sol})-(\ref{crci}) we get
\begin{eqnarray}\label{omegakink}
    \begin{aligned}
        &\omega_{-}=\sqrt{\frac{2\zeta}{\zeta+1}}~\omega_{A_{\rm i}}=\omega_{\rm kink},\\
        &\gamma_{-}=0,
    \end{aligned}
\end{eqnarray}
where $\omega_{\rm kink}$ is the kink mode frequency of an untwisted loop in the thin tube approximation (see e.g. Goossens et al. 2009).

Here, we are interested to obtain a minimum value for the twist parameter $\alpha=\alpha_{\rm min}$ which is required to excite the resonance. The exact value of $\alpha_{\rm min}$ is determined as the root of the following equation
\begin{eqnarray}\label{alphamin1}
    \omega_{-}|_{\alpha=\alpha_{\rm min}}=\omega_A(R)|_{\alpha=\alpha_{\rm min}}.
\end{eqnarray}
We solve this numerically and conclude that the result obtained for $\alpha_{\rm min}$ satisfies the following relation
\begin{eqnarray}\label{alphamin2}
   \omega_{-}|_{\alpha=\alpha_{\rm min}}\simeq\omega_{\rm kink}.
\end{eqnarray}
Equating right hand sides of Eqs. (\ref{alphamin1}) and (\ref{alphamin2}), one can get an approximate expression for $\alpha_{\rm min}$ as
\begin{eqnarray}\label{alphamin3}
    \alpha_{\rm min}\simeq \pi \left(\frac{R}{L}\right)\left(\frac{l}{m}\right)\left(\sqrt{\frac{2\zeta}{\zeta+1}}-1\right).
\end{eqnarray}
For instance, taking $\zeta=2$ and $R/L=0.01$, for the fundamental ($l=1$), first overtone ($l=2$) and second overtone ($l=3$) kink ($m=1$) modes, Eq. (\ref{alphamin3}) gives $\alpha_{\min}=$0.0049, 0.0097 and 0.0146, respectively. The relative error of the results of Eq. (\ref{alphamin3}) with respect to the exact solutions of Eq. (\ref{alphamin1}) is about $0.6 \%$.

\begin{figure}
\centering
    \includegraphics[width=0.48\textwidth]{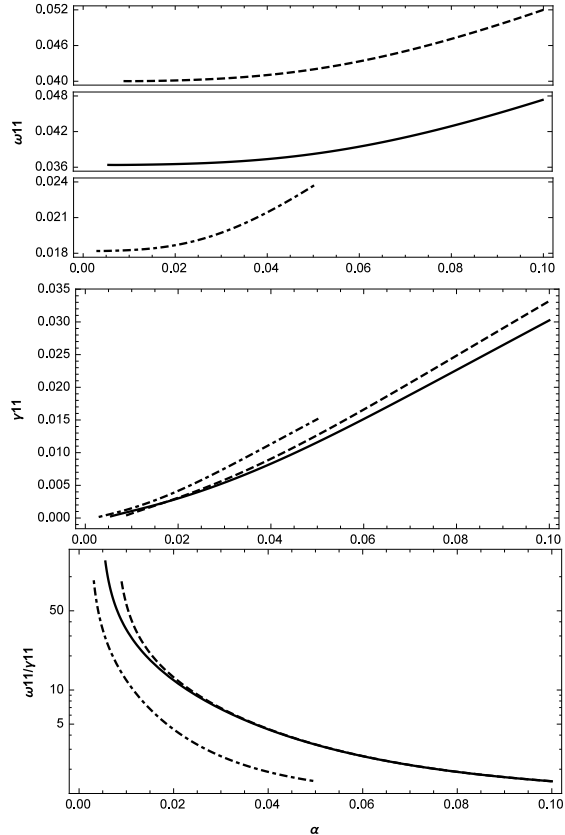}
    \caption[] {Frequency of the fundamental ($l=1$) kink ($m=1$) mode (top),
its damping rate (middle) and the ratio of the oscillation frequency to the
damping rate (Bottom) versus the twist parameter $\alpha$. Solid line: $\zeta=2, R/L=0.01$; Dashed line: $\zeta=4, R/L=0.01$; Dash-Dotted line: $\zeta=2, R/L=0.005$. Other auxiliary parameters as in Fig. \ref{background}. Both
frequencies and damping rates are in units of $v_{A_{\rm i}}/R=2$
rad $\rm s^{-1}$.}
    \label{111}
 \end{figure}

\begin{figure}
\centering
    \includegraphics[width=0.48\textwidth]{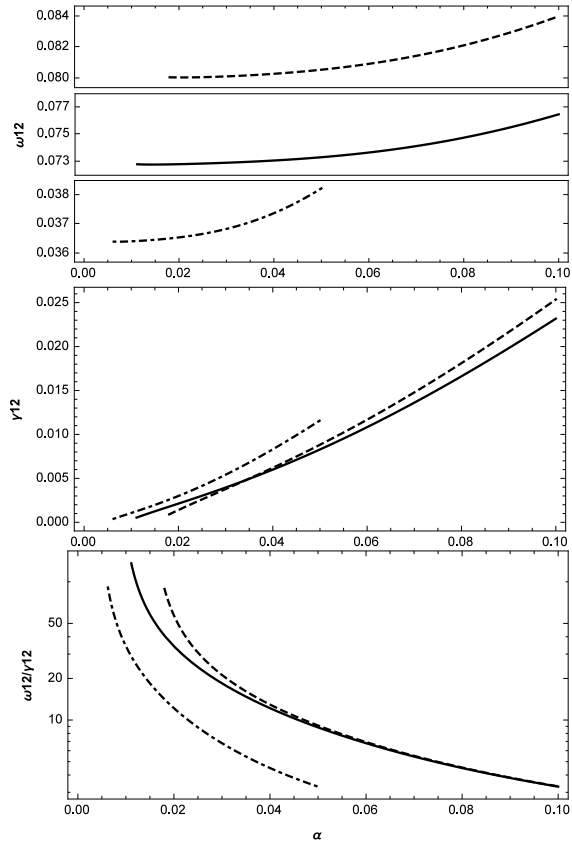}
    \caption[] { Same as Fig. \ref{111} but for the first overtone
($l=2$) kink ($m=1$) mode.}
    \label{112}
 \end{figure}

\begin{figure}
\centering
    \includegraphics[width=0.48\textwidth]{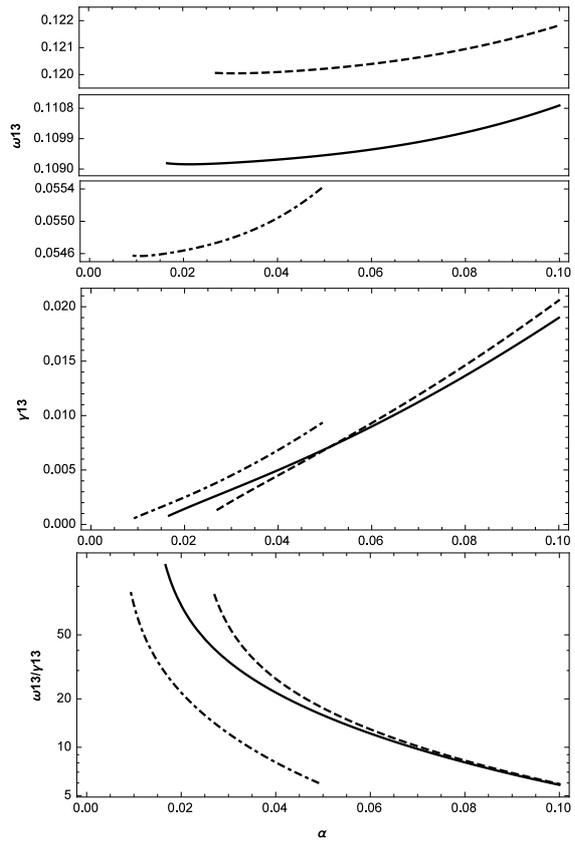}
    \caption[] {Same as Fig. \ref{111} but for the second overtone
($l=3$) kink ($m=1$) mode.}
    \label{113}
 \end{figure}

Figures \ref{111}, \ref{112}, and \ref{113} illustrate the frequencies ($\omega_{-}=\omega_{ml}$), the damping rates ($\gamma_{-}=\gamma_{ml}$) and the ratio of the frequency to
the damping rate $(\omega_{ml}/\gamma_{ml})$ of the fundamental ($l=1$) and first/second overtone $(l=2,3)$ kink ($m=1$) modes versus the twist
parameter $\alpha$, respectively. Figures show that (i) the
frequencies and damping rates increase when the twist parameter increases. (ii) The ratio of the
oscillation frequency to the damping rate $\omega/\gamma$ decreases when the twist parameter increases. The result, interestingly enough, is that for the fundamental ($l=1$) kind ($m=1$) mode, for the twist parameter $\alpha=0.0147$ we obtain $\omega_{11}/(2\pi \gamma_{11})=3$ which is in good agreement with the observations reported by Nakariakov et al. (1999), Wang \& Solanki (2004), and Verwichte et al. (2004). Here, the behaviour of $\omega$ and $\gamma$ versus  the twist parameter $\alpha$ are the same as that obtained by Karami \& Bahari (2010) but for the resonant absorption due to the radial density structuring. (iii) For a given $\alpha$ and $\zeta$, when $R/L$ increases then $\omega$ increases, $\gamma$ decreases and $\omega/\gamma$ increases. (iv) For a given $\alpha$ and $R/L$, when $\zeta$ increases, the frequency and the ratio $\omega/\gamma$ increase. However, this behavior for the damping rate holds only for large values of the twist parameter. Moreover, increasing $\zeta$ does not affect the ratio $\omega/\gamma$ for large values of $\alpha$. (v) Note that based on Eq. (\ref{alphamin3}), $\alpha_{\rm min}$ is a function of $m$, $l$, $\zeta$ and $R/L$. Therefore, for each set of these parameters, the start points of the diagrams in Figs. \ref{111} to \ref{113} have different $\alpha_{\rm min}$ values. Also to avoid the kink instability in our model, following Hood \& Priest (1979) we consider the twist value $\phi_{\rm twist}\lesssim \phi_{c\rm}=3.3\pi$ and consequently obtain an upper limit for the twist parameter as $\alpha_{\rm max}=(R/L)\phi_{\rm c}$. Hence, the diagrams in Figs. \ref{111} to \ref{113} with different $R/L$ have different $\alpha_{\rm max}$ cut-offs.

\begin{figure}
\centering
    \includegraphics[width=0.48\textwidth]{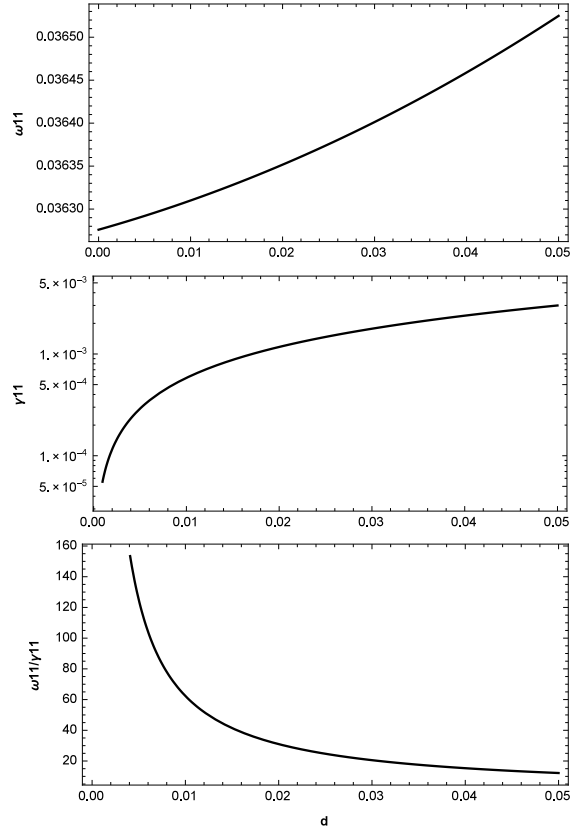}
    \caption[] {Frequency of the fundamental ($l=1$) kink ($m=1$) mode (top),
its damping rate (middle) and the ratio of the oscillation frequency to the
damping rate (bottom) for $\alpha=0.05$ versus the thickness of the twisted layer, $d$ (in units of $R=1000$ km). Other auxiliary parameters
        as in Fig. \ref{background}.} \label{111p}
 \end{figure}

Figure \ref{111p} presents variations of $\omega$, $\gamma$ and $\omega/\gamma$ for the fundamental ($l=1$) kink ($m=1$) mode versus the thickness of the twisted layer, $d\equiv R-a$. Figure shows that when $d$ increases both frequency and damping rate increase but the ratio of the frequency to the damping rate decreases. This behaviour holds also for the first and second overtone ($l=2,3$) kink modes. Figure \ref{111p} clarifies that in the limit of $d\rightarrow 0$ (i.e. $a\rightarrow R$), we have $\omega_{11}\rightarrow \omega_{\rm kink}=\sqrt{\frac{2\zeta}{\zeta+1}}~\omega_{A_{\rm i}}\simeq0.03628$ (in units of 2 rad s$^{-1}$) and $\gamma_{11}\rightarrow 0$. Note that the result of $\omega/\gamma$ for the fundamental mode of kink oscillations is in good agreement with that obtained for the resonant absorbtion due to the radial density structuring (see e.g. Ruderman \& Roberts 2002; Ruderman \& Terradas 2013).


\begin{figure}
\centering
    \includegraphics[width=0.48\textwidth]{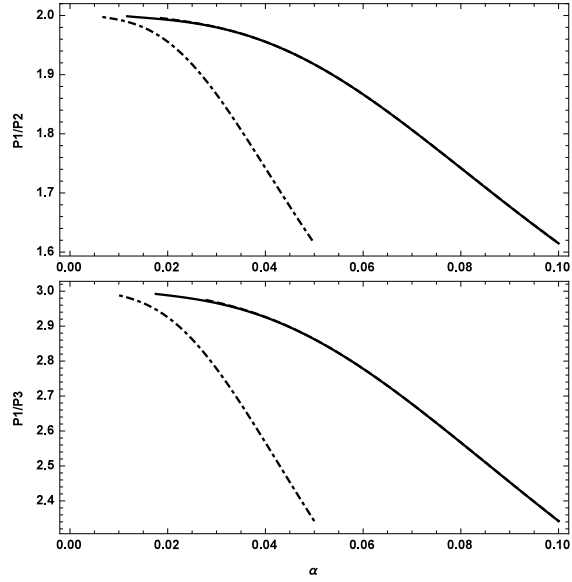}
    \caption[] {Period ratios of the fundamental to the
        first overtone, $P_1/P_2$, and to the second overtone, $P_1/P_3$,
        of the kink ($m=1$) modes versus the twist parameter $\alpha$. Solid line: $\zeta=2, R/L=0.01$. Dashed line (which overlaps the solid one): $\zeta=4, R/L=0.01$. Dash-Dotted line: $\zeta=2, R/L=0.005$. Other auxiliary parameters
        as in Fig. \ref{background}.} \label{pp}
 \end{figure}

 \begin{figure}
\centering
    \includegraphics[width=0.48\textwidth]{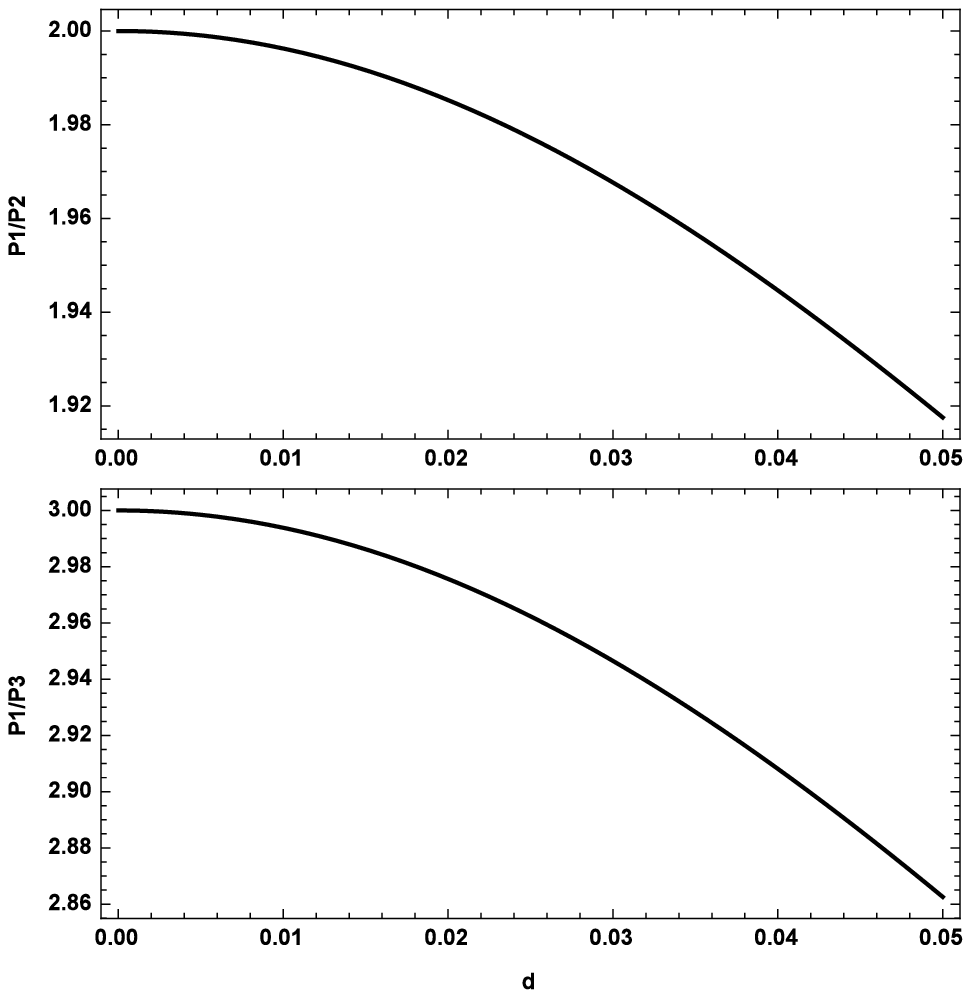}
    \caption[] {Period ratios of the fundamental to the
        first overtone, $P_1/P_2$, and to the second overtone, $P_1/P_3$,
        of the kink ($m=1$) modes for $\alpha=0.05$ versus the thickness of the twisted layer, $d$ (in units of $R=1000$ km). Other auxiliary parameters
        as in Fig. \ref{background}.} \label{ppd}
 \end{figure}

In Fig. \ref{pp}, the period ratios of the fundamental to
the first overtone, $P_1/P_2$, and to the second overtone,
$P_1/P_3$, of the kink ($m=1$) modes are plotted versus the twist
parameter $\alpha$. Figure shows that (i) when the twist parameter increases, the
values of $P_1/P_2$ and $P_1/P_3$ decrease from their canonical
values, 2 and 3, respectively. (ii) The density ratio $\zeta=\rho_{\rm i}/\rho_{\rm e}$ does not affect the period ratio. Notice that the solid curve overlaps the dashed one. (iii) For a given $\alpha$ and $\zeta$, the period ratio decreases when $R/L$ decreases. In Fig. \ref{ppd}, we plot variations of the period ratios $P_1/P_2$ and $P_1/P_3$ versus the thickness of the twisted layer $d$. Figure \ref{ppd} shows
that the period ratios decrease with increasing $d$.

Note that the deviations of the period ratios
$P_1/P_2$ and $P_1/P_3$ from their canonical values provide useful
information about the magnetic twist structuring within the loop.
For instance, the observed value $P_1/P_2=1.795$ reported by Van
Doorsselaere et al. (2007) can be justified with the twist parameter
$\alpha=0.0718$. Van Doorsselaere et al. (2007) also reported the identification of the second overtone of the kink mode by re-analyzing the transverse oscillations of coronal loops observed by TRACE on 13 May 2001. The period ratio of fundamental to second overtone
kink mode found to be $P_1/P_3=2.89$ which can be justified with the
twist parameter $\alpha=0.0460$. Table \ref{table} summarizes the twist parameters and the corresponding twist values, $\phi_{\rm twist}=\big(\frac{L}{R}\big)\alpha$, predicted by our model in order to justify some observed period ratios $P_1/P_2$ and $P_1/P_3$ of
the kink ($m=1$) modes. Note that the magnetic twist values $\phi_{\rm twist}$ predicted by our model are in the range of the observational values reported by Kwon \& Chae (2008) and Wang et al. (2015).

\begin{table} \centering\small{\caption{Magnetic
twist parameter $\alpha=B_\varphi/B_z$ and corresponding twist
value $\phi_{\rm twist}=\big(\frac{L}{R}\big)\alpha$ predicted by our model for some observational period ratios $P_1/P_2$ and $P_1/P_3$ of
the kink ($m=1$) modes. Other auxiliary parameters as in Fig. \ref{background}.}
\begin{tabular}{l c c | c | c}
\hline
\hline
 Reference & $P_1/P_2$ & $\alpha$ & $\phi_{\rm twist}/\pi$ \\
\hline
 Van Doorsselaere et al. (2007) & $1.795\pm0.051$  & 0.0718 & 2.28\\
 Van Doorsselaere et al. (2009) & $1.980\pm0.002$  & 0.0295 & 0.94\\
 Ballai et al. (2011)           & $1.82\pm0.02$  & 0.0678 & 2.16\\
\hline
\hline
 Reference & $P_1/P_3 $ & $\alpha$ & $\phi_{\rm twist}/\pi$ \\
\hline
  Van Doorsselaere et al. (2009) & $2.89\pm0.14$ & 0.0460  & 1.46\\
\hline
\end{tabular}
\label{table}}
\end{table}
\section{Conclusions}\label{sec6}

Here, we investigated the resonant absorption of kink MHD
modes by magnetic twist in coronal loops. To this
aim, we considered a thin straight cylindrical flux tube with a twisted magnetic field in a thin layer at the boundary of the loop
and a straight magnetic field everywhere else. The magnetic twist causes a radial Alfv\'{e}n
frequency gradient and consequently gives rise to the
resonant absorption. We assumed the plasma density to be constant but different in the interior and exterior regions of the loop.
We obtained the solutions of ideal MHD equations for the
interior and exterior regions of the tube. Then, we investigated a number of possible instabilities that may arise in our model and concluded that these instabilities can be avoided in the present model. We also derived the
dispersion relation by using the appropriate connection formula
introduced by Sakurai et al. (1991a). In thin tube thin boundary approximation, we solved analytically the dispersion
relation and obtained the frequencies and
damping rates of the fundamental $(l=1)$ and first/second overtone $(l=2,3)$
kink $(m=1)$ modes. Our results show the following.

 \begin{itemize}
\item The frequencies and damping rates increase when the twist parameter increases.

\item The ratio of the fundamental frequency to its corresponding damping rate of the kink ($m=1$) modes can well justify the rapid damping of kink MHD waves ($\omega/(2\pi \gamma)\simeq 3$) reported by the observations. This confirms the high efficiency of resonant absorbtion due to the magnetic twist.

\item For a given twist parameter $\alpha$ and density ratio $\zeta=\rho_{\rm i}/\rho_{\rm e}$ by increasing $R/L$, the frequencies increase, the damping rates decrease and the ratio of the frequency to the damping rate increases.

\item By increasing the thickness of the twisted layer, the frequency and the damping rate increase but the ratio of the frequency to the damping rate and the period ratio decrease.

\item For a given $\alpha$ and $\zeta$, the period ratio decreases with decreasing $R/L$. Furthermore, the density ratio $\zeta$ does not affect the period ratio.

\item The period ratios $P_1/P_2$ and $P_1/P_3$ with increasing the twist parameter, decrease from 2 and 3, respectively. For some special values of the twist parameter, the values of $P_1/P_2$ and $P_1/P_3$ predicted by our model can justify the observations.

 \end{itemize}

\section*{Acknowledgments}
The authors thank the anonymous referee for very valuable comments. The authors
also thank Anna Tenerani, Michael Ruderman, Robert Erd\'{e}lyi and Markus Aschwanden for useful discussions.


\end{document}